\documentstyle[epsf,twoside,fleqn,espcrc2]{article}

\title{The glueball spectrum from novel improved actions.}

\author{
        Colin Morningstar\address{Dept. of Physics, University of California at San Diego,
        La Jolla, California 92093-0319, USA}
        and 
        Mike Peardon\thanks{Poster presented by M.P.}
        \address{NIC, Forschungszentrum J\"ulich, J\"ulich D-52425, Germany}}
       
\begin{document}

\begin{abstract}
Results for the inter-quark potential and low-lying $SU(3)$ glueball spectrum from simulations using 
a new improved action are presented. The action, suitable for highly anisotropic
lattices, contains a two-plaquette term coupling with a negative coefficient as 
well as incorporating Symanzik improvement. 
\end{abstract}

\maketitle

\section{INTRODUCTION}

 The QCD glueball spectrum has been investigated in low-cost simulations using anisotropic lattices
\cite{glueballs_su3}.
To reduce the computational overhead, the spatial lattice was kept rather 
coarse (0.2-0.4 fm) while the temporal spacing was made much finer. The fine 
temporal grid allows 
adequate resolution of the Euclidean-time decay of appropriate correlation functions which, for 
gluonic states are rather noisy and fall too rapidly on coarse lattices.

In these simulations, the scalar glueball suffered from large finite cut-off effects. The
mass in units of $r_0$ fell sharply until the spatial lattice spacing, $a_s$ was about 0.25 fm when
the mass rose again; the ``scalar dip''.

At the conference last year, we presented results from simulations with an anisotropic Wilson
``two-plaquette'' action which included a term constructed
from the product of two parallel plaquettes on adjacent time-slices \cite{in_search}. This was found
to reduce the scalar dip significantly. Here, we report on the status of 
simulations in progress using a Symanzik-improved action including a similar two-plaquette term.

In this study, we tune the anisotropy parameter in the lattice action to recover Euclidean
invariance in the ``sideways'' potential. With these parameters fixed, we investigate the inter-quark
potential for this action as an initial test that the benefits of the Symanzik program are preserved
by the addition of the extra term. We are currently computing the glueball spectrum for
this action. 

\section{THE ACTION}

Following Ref. \cite{in_search}, we begin with the plaquette operator,

\[
P_{\!\mu\nu}(x) = \frac{1}{N} \mbox{ReTr } U_\mu(x) U_\nu(x\!+\!\hat{\mu}) 
                   U^\dagger_\mu(x\!+\!\hat{\nu}) U^\dagger_\nu(x).
\]
The Wilson (unimproved) discretisation of the magnetic field strength is then constructed from the
spatial plaquette.
\begin{eqnarray}
\Omega_s & = & \sum_{x,i>j} \left\{ 1 - P_{ij}(x) \right\} \nonumber \\
         & = & \frac{\xi_0}{\beta} \int\!\! d^4\!x \mbox{ Tr } B^2 + {\cal O}(a_s^2),
\end{eqnarray}
where $i,j$ are spatial indices and $\xi_0$ is the anisotropy, $a_s/a_t$ at tree-level in 
perturbation theory.  We introduce a term which correlates pairs of spatial plaquettes separated by 
one site temporally 
\begin{equation}
\Omega_{s}^{(2t)} = \frac{1}{2}\sum_{x,i>j} 
           \left\{ 1 - P_{ij}(x) P_{ij}(x+\hat{t})\right\}.
\end{equation}

The separation of the two plaquettes allows the standard Cabibbo-Marinari and 
over-relaxation gauge-field update methods to be applied. Including
two-plaquette terms adds a computational overhead of only $10\%$ to our improved
action workstation codes. 

It can be shown that for all $\omega$, the operator combination,
\begin{equation}
\tilde{\Omega}_s = (1+\omega) \; \Omega_s - \omega \; \Omega_{s}^{(2t)}
\label{eqn:twoplaq}
\end{equation}
has an identical expansion in powers of $a_{s,t}$ (at tree-level) to $\Omega_s$ up to 
${\cal O}(a_s^4)$. Thus, starting from the improved action $S_{I\!I}$ used in Refs. 
\cite{glueballs_su3,glueballs_su2}, it is straightforward to construct a Symanzik improved,
two-plaquette action by simply replacing the spatial plaquette term in 
$S_{I\!I}$ with the linear
combination $\tilde{\Omega}_s$ of Eqn. \ref{eqn:twoplaq}. In full, this action is 
\begin{eqnarray}
    && S_\omega  =   \nonumber \\
    &&   \frac{\beta}{\xi_0} \left\{
    \frac{5(1+\omega)}{3 u_s^4} \Omega_s
  - \frac{5\omega}{3 u_s^8}     \Omega_{s}^{(2t)}
  - \frac{1}{12 u_s^6}          \Omega^{(R)}_{s}
         \right\} \nonumber \\
    &&      +          
        \beta \xi_0 \left\{ 
    \frac{4}{3 u_s^2  u_t^2} \Omega_t
  - \frac{1}{12 u_s^4 u_t^2} \Omega^{(R)}_{t}
         \right\},  \label{eqn:twoplaq-action}
\end{eqnarray}
with $\Omega_t$ the temporal plaquette and $\Omega^{(R)}_{s}$,$\Omega^{(R)}_{t}$ the $2\times1$ 
rectangle in the $(i,j)$ and $(i,t)$ planes respectively. This action has leading 
${\cal O}(a_s^4,a_t^2, \alpha_s a_s^2)$ discretisation errors and only 
connects sites on adjacent time-slices, ensuring the free gluon propagator has 
only one real mode.

The free parameter $\omega$ is chosen such that the approach to the QCD continuum is made on a
trajectory far away from the critical point in the plane of fundamental-adjoint couplings. 
Close to the QCD fixed point, physical quantities should be weakly dependent on $\omega$. This
provides us with a consistency check, however the data presented here are for one value only, 
$\omega=3$.

\begin{figure}[t]
\setlength{\epsfxsize}{7.5cm}
\epsfbox{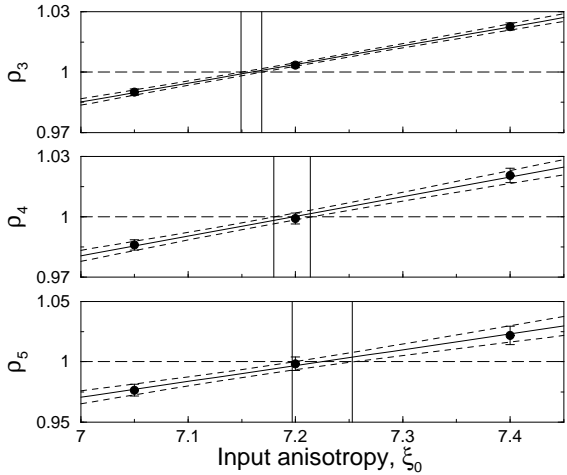}
\vspace{-6ex}
\caption{The ratio of Eqn. \protect{\ref{eqn:rho-def}} for $n=3,4,5$
\label{fig:xi_bare}}
\end{figure}

\section{TUNING THE ANISOTROPY}

\begin{figure}[t]
\setlength{\epsfxsize}{7.5cm}
\epsfbox{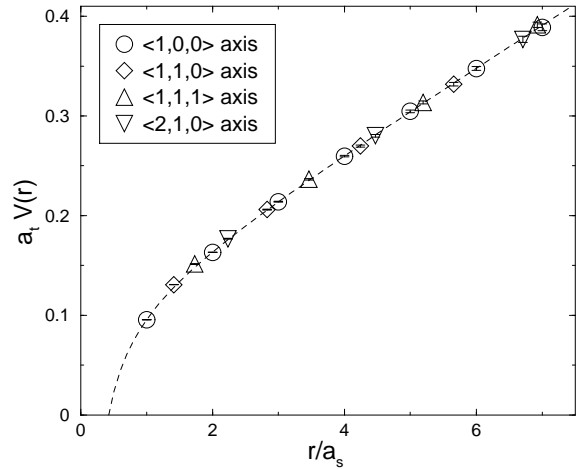}
\vspace{-6ex}
\caption{The inter-quark potential of $S_{I\!I}$ with $a_s\approx 0.22$ fm for sources separated 
         along different lattice axis. 
   \label{fig:V}}
\end{figure}

At finite coupling, the anisotropy measured using a physical probe differs from the parameter
in the action at ${\cal O}(\alpha_s)$ \cite{aniso}. In previous calculations, we relied 
upon the smallness of these renormalisations for the (plaquette mean-link improved) action 
$S_{I\!I}$. For the action of Eqn. \ref{eqn:twoplaq-action}, these renormalisations are larger and 
thus we chose to tune the input parameter in the action to ensure that the potentials measured 
along anisotropic axes matched. We follow a similar procedure to Ref. \cite{klassen}. The potentials
between two static sources propagating along the z-axis for separations on both fine and 
coarse axes, $V_t$ and $V_s$ respectively, are measured using smeared Wilson loops. Since the UV 
divergences due to the static sources are the same, tuning $\xi_0$ such that the ratio
\begin{equation}
  \rho_n = \frac{a_s V_s(n a_s)}{ a_s V_t (m n a_t)} \equiv 1,
  \label{eqn:rho-def}
\end{equation}
implies the anisotropy $\xi_V = m \; (m \in Z)$. A consistency check is provided by studying 
different coarse source separations, $n a_s$. Fig. \ref{fig:xi_bare} shows this tuning for 
$n=3,4,5$, where the desired anisotropy is 6. Consistency is observed for $n=4$ and 5 and the 
appropriate $\xi_0$ is found to better than $1\%$.

\section{SIMULATION RESULTS}

\subsection{The inter-quark potential}

  The replacement of the spatial plaquette in $S_{I\!I}$ with $\tilde{\Omega}_s$ of Eqn. 
\ref{eqn:twoplaq} should lead only to changes in the irrelevent operators responsible for 
${\cal O}(a_s^4, \alpha_s a_s^2)$ errors. To test this replacement still generates an
improved action with the good rotational invariance of $S_{I\!I}$, the inter-quark potential was
computed for a variety of different inter-quark lattice orientations. The potential is shown in
\ref{fig:V}, and shows excellent rotational invariance. We conclude that the benefits of the 
Symanzik improvement programme are preserved by including the two-plaquette term for a typical
value of $\omega$ useful for glueball simulation. 

\subsection{The glueball spectrum}

\begin{figure}[t]
\setlength{\epsfxsize}{7.5cm}
\epsfbox{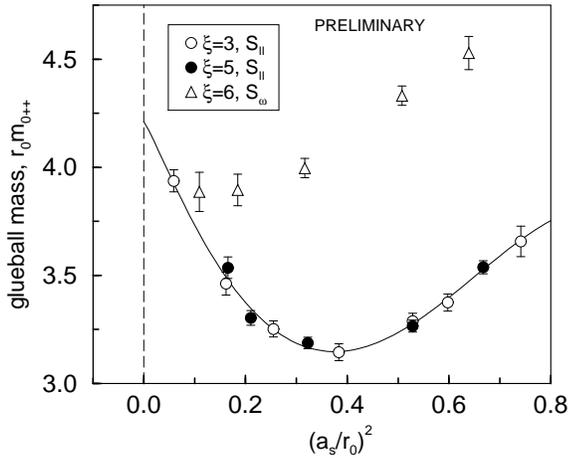}
\vspace{-6ex}
\caption{The scalar glueball of both Symanzik improved actions, $S_{I\!I}$ and $S_\omega$
   \label{fig:glue-1}}
\end{figure}

\begin{figure}[t]
\setlength{\epsfxsize}{7.5cm}
\epsfbox{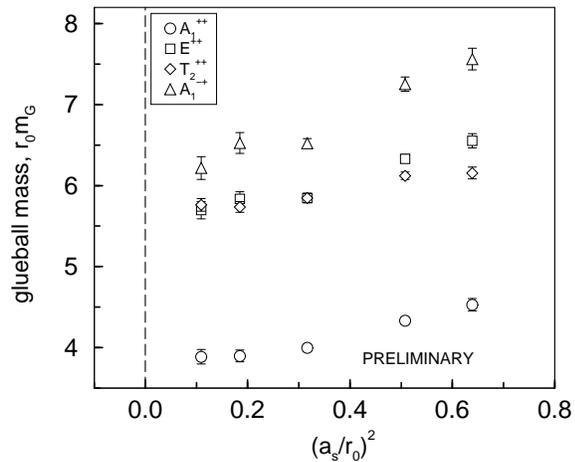}
\vspace{-6ex}
\caption{The scalar, tensor and pseudoscalar glueballs of $S_\omega$
   \label{fig:glue-2}}
\end{figure}

At present, we are computing the glueball spectrum on the $\xi_V=6$ tuned lattices. Preliminary data
are presented in Figs. \ref{fig:glue-1} and \ref{fig:glue-2}. 
In Fig. \ref{fig:glue-1}, the finite-lattice-spacing artefacts in the scalar glueball mass for the 
new action are compared to those of $S_{I\!I}$. The lattice cut-off dependence is seen to be
significantly reduced and for the range of lattice spacings studied here, the 
mass rises monotonically with lattice spacing rather than
falling first to a minimum. 
Fig. \ref{fig:glue-2} shows the lattice spacing dependence on the tensor 
 and pseudoscalar glueballs. Their lattice spacing 
dependence is similar to the form for $S_{I\!I}$ and consistent with leading ${\cal O}(a_s^4)$ behaviour.

\section{CONCLUSIONS}

Preliminary data from our simulations of the Symanzik improved action of Eqn. 
\ref{eqn:twoplaq-action} suggest the scalar dip is removed by inclusion of a two-plaquette term with
negative coefficient, consistent with the argument that the poor scaling of the scalar glueball,
even after Symanzik improvement, is caused by the presence of a nearby critical point. 

The inter-quark potential on this new action exhibits equally good rotational symmetry to the 
improved actions of Refs. \cite{glueballs_su3,glueballs_su2}.

\end{document}